\documentstyle[psbox,graphicx,epstopdf]{WORKSHOP}

\begin{document}

\title{3D Hydrodynamic \& Radiative Transfer Models of \\X-ray Emission from Colliding Wind Binaries}

\author{Christopher M.~P. Russell$^1$, Atsuo T. Okazaki$^1$, Stanley P. Owocki$^2$,\\
Michael F. Corcoran$^3$, Kenji Hamaguchi$^3$ and Yasuharu Sugawara$^4$
\\[12pt]
$^1$  Hokkai-Gakuen University, Sapporo, Japan \\
$^2$  University of Delaware, Newark, DE, USA \\
$^3$  NASA/GSFC, Greenbelt, MD, USA \\
$^4$  Chuo University, Tokyo, Japan \\
{\it E-mail (CMPR): crussell@udel.edu}}

\abst{Colliding wind binaries (CWBs) are unique laboratories for X-ray astrophysics.  The massive stars in these systems possess powerful stellar winds with speeds up to $\sim$3000 km\,s$^{-1}$, and their collision leads to hot plasma (up to $\sim10^8$K) that emit thermal X-rays (up to $\sim$10 keV).  Many X-ray telescopes have observed CWBs, including \textit{Suzaku}, and our work aims to model these X-ray observations. 
We use 3D smoothed particle hydrodynamics (SPH) to model the wind-wind interaction, and then perform 3D radiative transfer to compute the emergent X-ray flux, which is folded through X-ray telescopes' response functions to compare directly with observations.  In these proceedings, we present our models of \textit{Suzaku} observations of the multi-year-period, highly eccentric systems $\eta$ Carinae and WR\,140.  The models reproduce the observations well away from periastron passage, but only $\eta$ Carinae's X-ray spectrum is reproduced at periastron; the WR\,140 model produces too much flux during this more complicated phase.}

\kword{hydrodynamics, radiative transfer, X-rays, massive stars, binaries, eta Carinae, WR 140}

\maketitle
\thispagestyle{empty}

\section{Smoothed Particle Hydrodynamics}

The wind-wind interaction of CWBs is modeled with an SPH code (Okazaki et al. 2008) that launches SPH particles from two point masses; these wind particles are accelerated according to a $\beta$=1 velocity law.  The calculation includes radiative cooling, which is implemented via the exact integration scheme (Townsend 2009), and uses a cooling function from \texttt{Cloudy} (Smith et al.\ 2008).  Solar abundances are assumed throughout.

\begin{figure*}[t]
\centering
  \includegraphics[height=60mm]{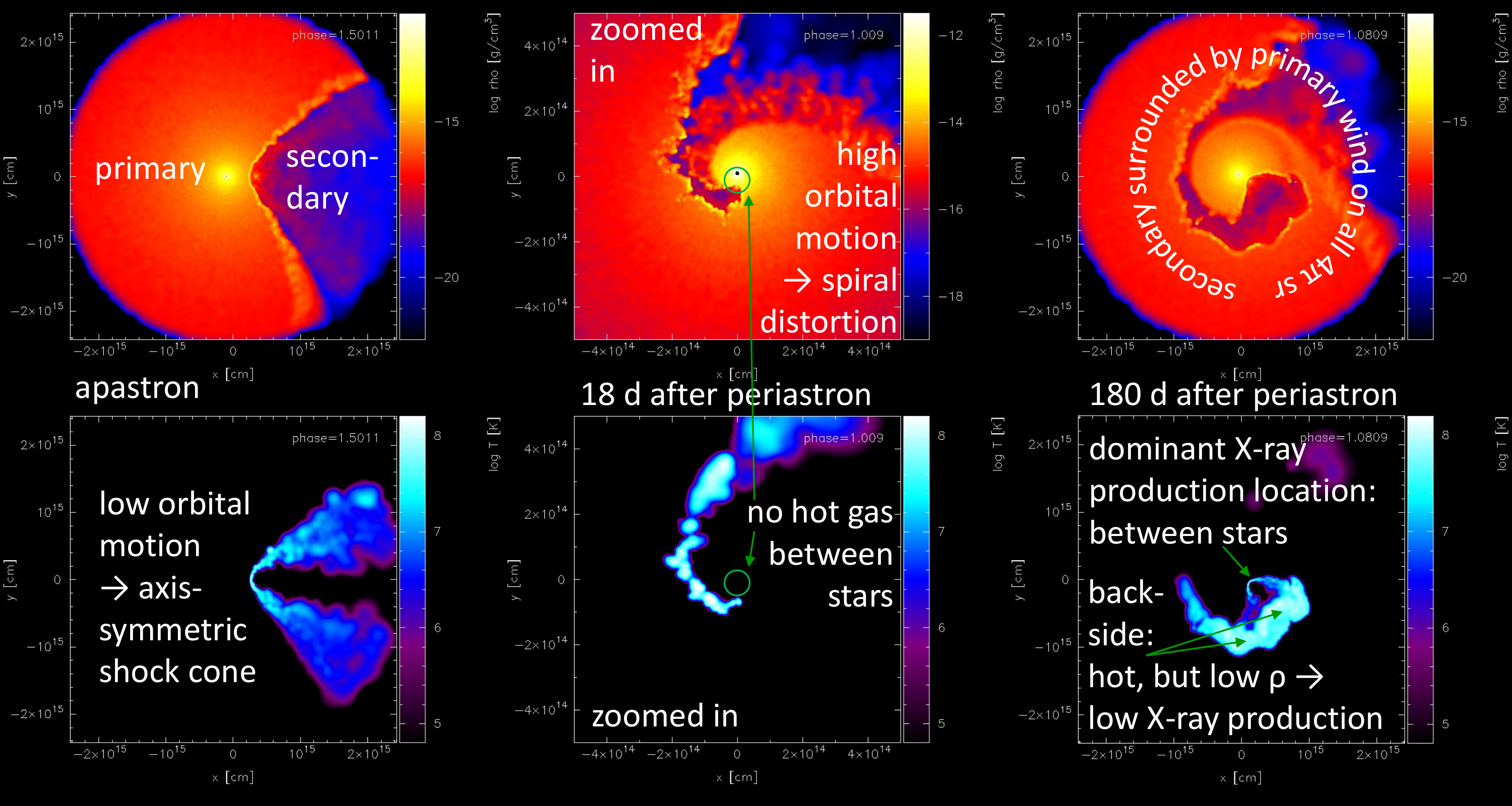}
  \caption{Density (top) and temperature (bottom) in the orbital plane of $\eta$ Carinae at three phases.  These hydrodynamic quantities vary significantly throughout the orbit, indicating that the model X-ray emission will as well.}
  \label{fi:SPH}
\end{figure*}

Figure \ref{fi:SPH} shows the density (top row) and temperature (bottom row) structure in the orbital plane of a 3D SPH simulation of $\eta$ Carinae (period $P=5.54$\,yr, eccentricity $e=0.9$).  The left panels show the axis-symmetric shock cone at apastron, which stems from the low orbital motion.  Near periastron, however, significant spiral distortion occurs, which necessitates the use of 3D.  Interestingly the hot gas between the stars, which is normally the site of the maximum X-ray emission since the collision is head on and the densities are highest, disappears around periastron.  Two effects cause this; the primary wind encroaches into the acceleration region of the secondary wind, so the shock velocity is not as high, and the shock switches from adiabatically cooling to radiatively cooling.  The steep dependence of the shock velocity on whether a shock in a CWB is adiabatic or radiative ($\sim$$v^4$, Stevens et al.\ 1992) means this transition happens quickly.  After periastron the secondary wind collides with primary wind over all solid angles, but the dominant X-ray emission comes from between the stars.

WR\,140 ($P=7.9$\,yr, $e=0.881$) shows similar behavior to $\eta$ Carinae, except that the spiral distortion around periastron recovers much quicker since the terminal speeds of the two winds are more similar.  Also the stellar radii are smaller, so the winds always collide at or near terminal speed, and thus there is always hot gas between the stars.  See Russell (2013) for more details on the SPH calculations and the full parameters for each CWB.

\section{X-ray Radiative Transfer}

The model X-ray flux is computed directly from the SPH simulations by solving the formal solution of radiative transfer, which we do by modifying the SPH visualization program \texttt{Splash} (Price 2007).  The emissivity (solar composition is assumed except for the WC wind in WR\,140) is from \texttt{APEC} (Smith et al.\ 2001), which is obtained via \texttt{XSpec} (Arnaud 1996), the circumstellar absorption is from \texttt{windtabs} (Leutenegger et al.\ 2010), and the ISM absorption is from \texttt{TBabs} (Wilms et al.\ 2000).  The radiative transfer is performed along a series of 1D rays through the SPH simulation; this creates a pixel map that is then summed to determine the flux at a particular energy.  The radiative transfer calculation is looped over energy to compute a spectrum, and this is finally folded through the response function of an X-ray instrument to directly compare the model spectra with the data in absolute units.

\begin{figure}
  \includegraphics[height=45mm]{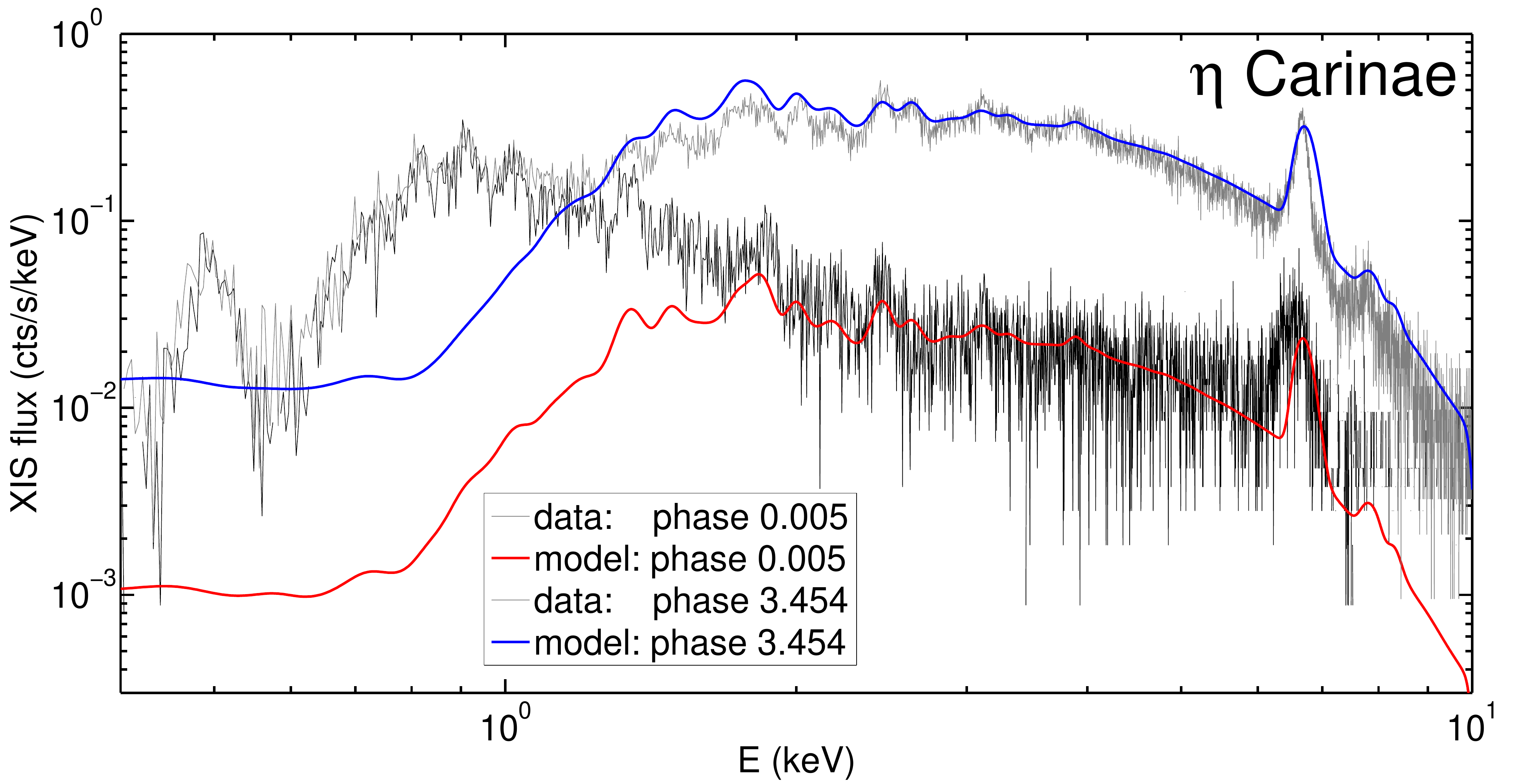}%
  \caption{\textit{Suzaku} spectra of $\eta$ Carinae at apastron (blue/gray) and periastron (red/black).  The soft X-ray flux of $\eta$ Carinae comes from a large region far outside the central colliding wind region (note that it does not change as a function of phase), so it is beyond the scope of our modeling.}
  \label{fi:XeC}
\end{figure}

\begin{figure}
  \includegraphics[height=45mm]{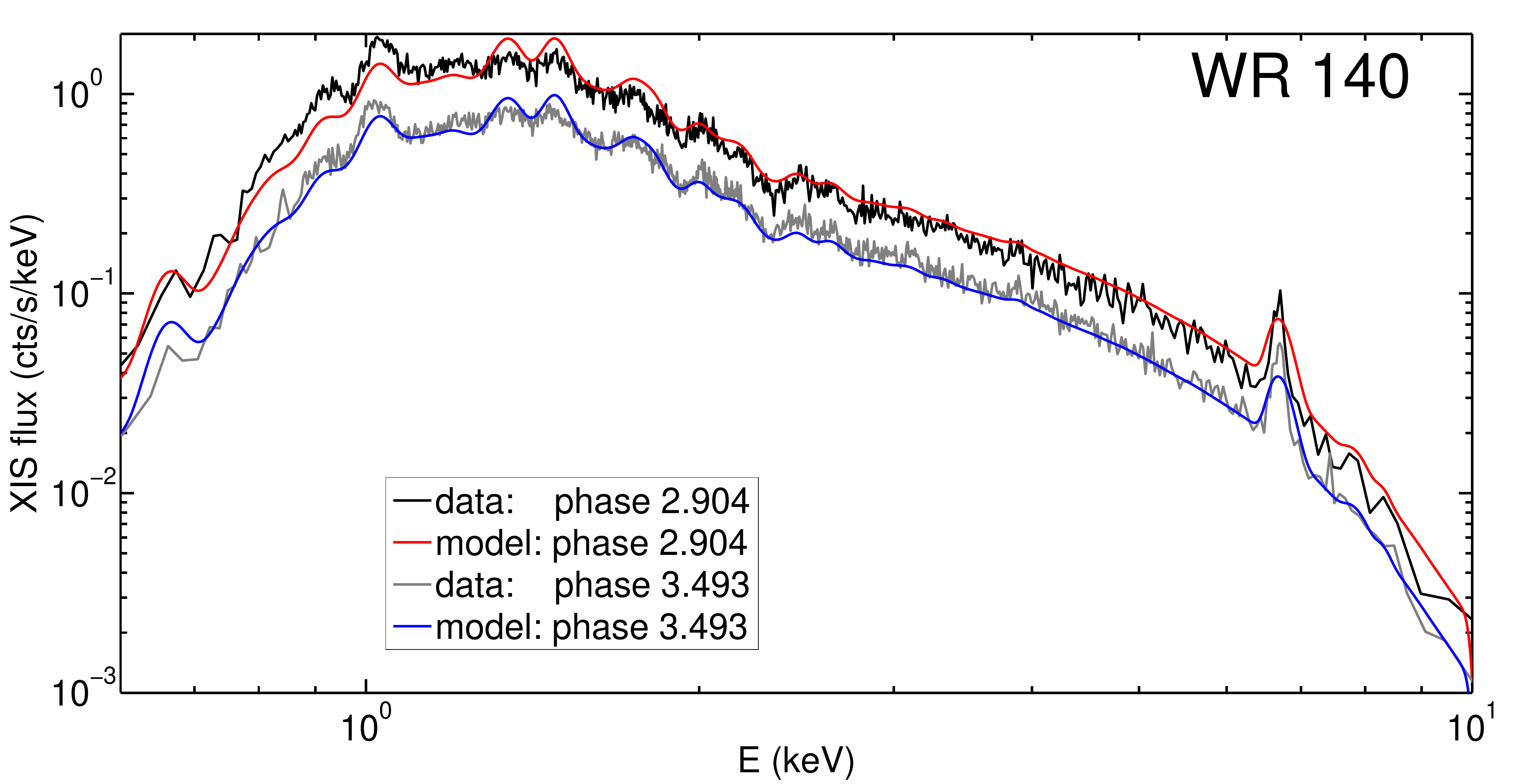}
  \caption{\textit{Suzaku} spectra of WR\,140 at apastron (blue/gray) and shortly before periastron (red/black).  Data from Sugawara et al.\ (2011).}
  \label{fi:XW}
\end{figure}

Figures \ref{fi:XeC} \& \ref{fi:XW} compare the model X-ray spectra of $\eta$ Carinae and WR\,140, folded through the \textit{Suzaku} XIS response function, with the data.  The agreement is good for all four observations shown, providing more evidence that the mass loss rate of $\eta$ Carinae's primary is $\sim$$10^{-3}M_\odot$\,yr$^{-1}$ (Hillier et al. 2001).  The model does not match the WR\,140 observations at periastron (not shown); the spectral shape is good, but the model flux level gets $\sim$3$\times$ too high at its worst disagreement, requiring further work to resolve this discrepancy.


\section*{References}

\re
Arnaud, K.~A.\ 1996, in ASP Conference Series, Vol. 101, ed. G.~H. Jacoby \& J. Barnes, 17

\re
Hillier, D.~J., Davidson, K., Ishibashi, K., \& Gull, T.\ 2001, ApJ, 553, 837

\re
Leutenegger, M.~A., Cohen, D.~H., \& Zsargó, J., et al.\ 2010, ApJ, 719, 1767

\re
Okazaki, A.~T., Owocki, S.~P., Russell, C.~M.~P., \& Corcoran, M.~F.\ 2008, MNRAS, 388, L39

\re
Price, D.~J.\ 2007, PASA, 24, 159

\re
Russell, C.~M.~P.\ 2013, PhD thesis, Univ.\ of Delaware

\re
Smith, R.~K., Brickhouse, N.~S., Liedahl, D.~A., \& Raymond, J.~C.\ 2001, ApJLett, 556, L91

\re
Smith, B., Sigurdsson, S., \& Abel, T.\ 2008, MNRAS, 385, 1443

\re
Stevens, I.~R., Blondin, J.~M., \& Pollock, A.~M.~T.\ 1992, ApJ, 386, 265

\re
Sugawara, Y., Maeda, Y., et al.\ 2011, Bulletin de la Societe Royale des Sciences de Liege, 80, 724

\re
Townsend R.~H.~D.\ 2009 ApJS., 181, 391

\re
Wilms, J., Allen, A., \& McCray, R.\ 2000, ApJ, 542, 914

\label{last}

\end{document}